\documentclass[prb,showpacs,preprintnumbers,amsfonts,amssymb,amsmath,floats,twocolumn,aps]{revtex4}
 
\usepackage[pdftex]{graphicx}
\usepackage[pdftex,colorlinks=true,linkcolor=blue,citecolor=blue,filecolor=blue,baseurl=empty]{hyperref}
\usepackage{dcolumn}
\usepackage{bm}
\usepackage{color}
\usepackage{amsmath}

\newcommand{\ca}[1]{{\cal #1}}
\newcommand{\be}{\begin{equation}}
\newcommand{\ee}{\end{equation}}
\newcommand{\lr}[1]{\left(#1\right)}
\newcommand{\matrixel}[3]{\big< #1 \vphantom{#2#3} \big| #2 \big| #3 \vphantom{#1#2} \big>} % for Dirac matrix elements
\newcommand{\ket}[1]{\big| #1 \big>} % for Dirac bras
\newcommand{\avg}[1]{\big< #1 \big>} % for average
\newcommand{\norm}[1]{\left|#1\right|^2}
\newcommand{\rec}[1]{\frac{1}{#1}}
\newcommand{\pd}[2]{\frac{\partial #1}{\partial #2}} % for partial derivatives

\begin{document}

\title{Filling-dependent doublon dynamics in the one-dimensional Hubbard model}

\author{Roman Rausch and Michael Potthoff}
\affiliation{I. Institute for Theoretical Physics, University of Hamburg, Jungiusstra\ss{}e 9, 20355 Hamburg, Germany}

\begin{abstract}
The fate of a local two-hole doublon excitation in the one-dimensional Fermi-Hubbard model is systematically studied for strong Hubbard interaction $U$ in the entire filling range using the density-matrix renormalization group (DMRG) and the Bethe ansatz. 
For strong $U$, two holes at the same site form a compound object whose decay is impeded by the lack of phase space.
Still, a partial decay is possible on an extremely short time scale where phase-space arguments do not yet apply. 
We argue that the initial decay and the resulting intermediate state are relevant for experiments performed with ultracold atoms loaded into an optical lattice as well as for (time-resolved) CVV Auger-electron spectroscopy. 
The detailed discussion comprises the mixed ballistic-diffusive real-time propagation of the doublon through the lattice, its partial decay on the short time scale as a function of filling and interaction strength, as well as the analysis of the decay products, which are metastable on the intermediate time scale that is numerically accessible and which show up in the two-hole excitation (Auger) spectrum.
The ambivalent role of singly occupied sites is key to understanding the doublon physics: 
For high fillings, ground-state configurations with single occupancies are recognized to strongly relax the kinematic constraints and to open up decay channels.
For fillings close to half filling, however, their presence actually blocks the doublon decay. 
Finally, the analysis of the continua in the two-hole spectrum excludes a picture where the doublon decays into unbound electron holes for generic fillings, different from the limiting case of the completely filled band.
We demonstrate that the decay products as well as the doublon propagation should rather be understood in terms of Bethe ansatz eigenstates.
\end{abstract}


% 71.10.Fd lattice fermion models (hubbard model, etc.)
% 67.85.Lm Degenerate Fermi gases
% 82.80.Pv Electron spectroscopy (X-ray photoelectron (XPS), Auger electron spectroscopy (AES), etc.)
% 78.47.D- time resolved spectroscopy (>1 psec)

\maketitle

%---------------
\section{Introduction}
\label{sec:intro}
%---------------

Over the last two decades, ultracold atomic gases in optical lattices have emerged as prototypes for studying time-dependent, nonequilibrium processes in systems of strongly correlated quantum degrees of freedom, controlling them with unprecedented precision. \cite{Gia04,LSA+07,BDZ08,Ess10}
Of particular interest is the behavior of double occupancies or ``doublons.'' 
In the case of a single-orbital lattice system with fermionic spin-$1/2$ particles, like the famous Hubbard model, 
\cite{Gut63,Hub63,Kan63}
these are two particles with opposite spin projections occupying the same lattice site. 
Even if the mutual interaction is repulsive ($U>0$), they can form a bound object whose decay is strongly suppressed due to the lack of phase space. 
\cite{WTL+06,Rosch_etal_2008,Strohmaier_etal_2010}
This is in turn is a result of energy and momentum conservation together with the fact that the spectrum of one-particle energies in these systems is bounded from above.

A simple way of observing this phenomenon in real time is to prepare a {\em single} doublon at a particular site of an otherwise empty lattice, to let the system evolve and to look at the double occupancy as a function of time. 
This setup is also straightforward to achieve experimentally by switching off a trapping potential. Numerically, it has been investigated for both fermions \cite{Kajala_etal_2011, Hofmann_Potthoff_2012} and bosons, \cite{WTL+06,Boschi_etal_2014} and constitutes a comparatively simple two-body problem.

For a lattice model at a finite particle density, however, the problem attains a true many-body character. 
Several setups are now conceivable: 
One could directly prepare an initial state with definite site occupations, e.g.\ with a certain number of doubly occupied sites in a central region, and look at the emerging wavefronts in the subsequent time evolution.
This can be achieved by sudden switches of external fields \cite{Ganahl_etal_2012} 
or of model parameters. \cite{Langer_etal_2012, Vidmar_etal_2013, Hauschild_etal_2015, dePaula_etal_2016} 
In what could be termed ``geometrical quench,'' for example, one switches on the nearest-neighbor hopping between different isolated lattice segments in their respective ground states, but with different total double occupancies.
Alternatively, one may prepare an excited initial state with a doublon by suddenly creating, at the same site, two particles (or holes) in the ground state. 
This has been done by removing a single boson at unit filling for the Bose-Hubbard model \cite{Andraschko_Sirker_2015} and, for the Fermi-Hubbard model, by creating a nearest-neighbor particle-hole excitation in the half-filled ground state. \cite{Al-Hassanieh_etal_2008} 
Another approach is to prepare a Gaussian wave packet of doublons, where a compromise is made between localization in real and momentum space. 
This idea has been pursued, for example, to study spin-charge separation in one spatial dimension.\cite{Ulbricht_Schmitteckert_2009, Al-Hassanieh_etal_2013, Moreno_etal_2013} 

Doublon decay has two fundamentally different aspects which must be distinguished carefully:
On the one hand, decay in a high-order scattering process with many constituents in the final state is possible even for strong $U$, i.e.\ consistent with the requirements of energy and momentum conservation. 
At the same time, however, this is a rather unlikely event. 
Thus, doublon dynamics must be watched on a long time scale to let the decay processes become effective.
Evidence suggests \cite{Strohmaier_etal_2010, CGK12} that the decay time is exponentially long in the interaction strength, $\tau\sim e^U$, reflecting the fact that, with increasing $U$, higher and higher scattering orders are needed to render the process possible.

On the other hand, for short times, the energy (and momentum) conservation argument that restricts the available phase space is actually invalidated by the uncertainty principle. 
On this short time scale, $\tau \approx \hbar/U$, doublon decay is possible even via low-order scattering processes.
It is therefore expected that there is an initial rapid decay of the doublon excitation which leaves the system with a reduced total double occupancy in an intermediate state. 
This state is established once the kinematic restrictions have become fully active.
Depending on $U$, however, the remaining double occupancy of the intermediate state can be still quite different from the ground-state or thermal value which, in a high-order process, is approached subsequently on the exponentially long time scale.

The long-time limit is important in the context of the thermalization of isolated quantum systems (in particular when a macroscopically large number of doublons is excited) and can be studied, for instance, with ultracold atoms trapped in optical lattices. \cite{Strohmaier_etal_2010}
It is less relevant, however, for condensed-matter systems, where doublons are formed by electrons in a narrow conduction band with strongly screened Coulomb interaction $U$. 
Here, a fast doublon decay is facilitated by the fact that the spectrum of electronic excitations is not bounded from above and by the presence of additional decay channels involving lattice degrees of freedom, for example. 

Doublon decay on the short time scale, typically in the femtosecond regime, and the stability of doublon excitations in the intermediate state is the key to understand line shapes in high-resolution CVV Auger-electron spectroscopy (AES): \cite{Pot01b} 
After the Auger process, two electron holes are left behind in the conduction band and form a doublon.
The corresponding bound state is reflected in the spectrum as a correlation satellite, whose position shifts linearly with the binding energy $U$. \cite{Cini_1977, Sawatzky_1977} 
Doublon decay on the short timescale, on the other hand, is reflected in the appearance of the so-called ``band-like part'' of scattering states, where the two electrons propagate independently through the lattice. We have recently shown how this phenomenology extends to the whole filling range, where more complex  bound states may be formed. \cite{Rausch_Potthoff_2016}

The two-hole spectral function (see Eq.\ (\ref{eq:A}) below) is related to the real-time dynamics of the initial doublon excitation on the lattice via a Fourier transform. 
Hence, the dynamics on the short to intermediate time scale determines the overall broad features of the spectrum, such as its separation into a correlation satellite and a band-like part, while the long-time behavior should determine the ``fine structure.'' 
A high spectral resolution, increasing exponentially with increasing $U$, would be necessary to uncover the long-time decay of the doublon in the spectrum.

The present work concentrates on the doublon decay on the short to intermediate time scale, on its relation with the two-hole spectroscopy as well as on the real-time propagation of an initial doublon excitation on the lattice.
We aim at a detailed and systematic study in the whole filling range for the one-dimensional Fermi Hubbard model.
Here, in one spatial dimension, we can profit from well-established and powerful methods such as the density-matrix renormalization group (DMRG) \cite{Sch11} and the Bethe ansatz (BA). \cite{Essler_2005, Solyom_2010}
Furthermore, the one-dimensional case is highly interesting physically due to the phenomenon of spin-charge separation:
Besides the doublon decay, there is a further decay of electrons into antiholons, which carry just the electrons' charge, and spinons, which carry just their spin. 
Correspondingly, {\em electron holes} decay into spinons and holons.
This decay takes place on a time scale which is roughly set by the (inverse) nearest-neighbor hopping $T$. 
\cite{Ulbricht_Schmitteckert_2009,Moreno_etal_2013,Al-Hassanieh_etal_2013}
Hence, those decay products are expected to show up in the two-hole spectrum. 

The paper is organized as follows:
Section \ref{sec:doublons} gives an introduction by discussing the two-body case. 
In Sec.\ \ref{sec:decay} we analyze the two-hole spectral function and identify the decay products of the doublon with the help of the Bethe ansatz. 
The real-time dynamics of the decay process is addressed in Sec.\ \ref{sec:dynamics}, and in Sec.\ \ref{sec:prop} we study the dynamics of the propagation of the surviving doublon. 
Conclusions are given in Sec.\ \ref{sec:summary}.

%---------------
\section{Doublons}
\label{sec:doublons}
%---------------

Using standard notations, the Hubbard model is given by
\be
H = -T \sum_{\langle ij \rangle , \sigma} \lr{c^{\dagger}_{i\sigma} c_{j\sigma} + \text{h.c.}} + U\sum_{i} n_{i\uparrow} n_{i\downarrow} \: . 
\label{eq:ham}
\ee
Here, $c^{\dagger}_{i\sigma}$ creates an electron at lattice site $i$ with spin projection $\sigma = \uparrow,\downarrow$, and $n_{i\sigma}=c^{\dagger}_{i\sigma}c_{i\sigma}$ is the occupancy-number operator. Summation over ordered pairs of nearest neighbors is denoted by $\langle ij \rangle$. 
$U$ is the strength of the Hubbard interaction, and $T$ is the nearest-neighbor hopping amplitude which is set to unity to fix the energy and time scales ($\hbar = 1$).
We consider the model on the one-dimensional lattice and for band filling $n=N/L$, where $N$ is the total particle number, and $L$ is the number of lattice sites. 

\begin{figure}
\includegraphics[width=0.95\columnwidth]{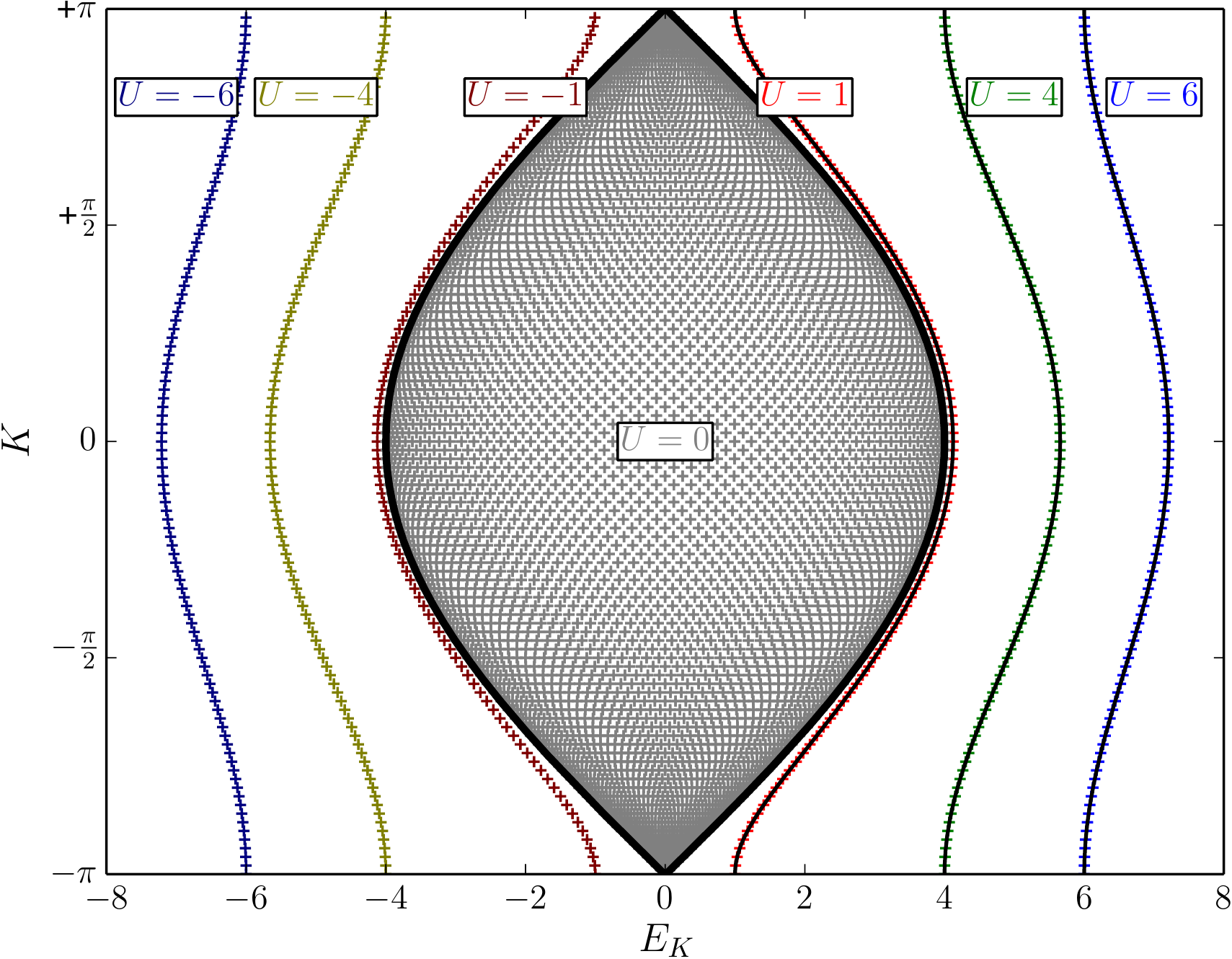}
\caption{Eigenvalues of the one-dimensional Hubbard model on $L=100$ lattice sites (periodic boundary conditions) in the subspace of two electrons spanned by the states given in Eq.\ (\ref{eq:doublonBasis}) and classified according to the center-of-mass momentum $K$. 
Calculations for different $U$ as indicated.
For $\big|U\big|>0$, only the bound-state solutions are shown.
}
\label{fig:doublonEigenvaluesBA}
\end{figure}

To start the discussion of doublon physics, it is instructive to consider simple limits, such as the case of two electrons with opposite spin directions in an otherwise empty lattice. 
The two-electron subspace is spanned by the $L^2$ states
\be
\ket{ij} = c^{\dagger}_{i\downarrow} c^{\dagger}_{j\uparrow} \ket{0} \: .
\label{eq:doublonBasis}
\ee
The problem can be reduced to a diagonalization of an $L\times L$ matrix by introducing relative and center-of-mass coordinates, i.e.\ $r=i-j$ and $R=\lr{i+j}/2$. \cite{Valiente_Petrosyan_2008, Valiente_Petrosyan_2009, Qin_etal_2014} 
Therewith, the eigenstates can be classified according to the center-of-mass momentum $K$ and fall into two categories: Scattering states, where the electrons propagate independently of each other, and bound states, where the wave function falls off exponentially with $r$ and which are associated with doubly occupied sites.
This is demonstrated by the results of a simple exact calculation shown in Fig.\ \ref{fig:doublonEigenvaluesBA}. 
Bound states appear for both attractive ($U<0$) and repulsive ($U>0$) interaction and are separated in total energy and momentum from the two-particle continuum.

The dispersion of the bound states can be understood analytically for strong $U$: 
In the limit $U \gg T$ and neglecting single occupancies, an effective doublon model is obtained by means of the Schrieffer-Wolff transformation: \cite{Fazekas_1999,Auerbach_2012,Rosch_etal_2008}
\begin{eqnarray}
  H_{\text{eff}} 
  &=& 
  \frac{J}{2} \sum_{\langle ij \rangle} \lr{d^\dagger_i d_j + \mbox{h.c.}} 
  \nonumber \\
  &+& 
  \lr{J+U} \sum_i n_i^d 
  - 
  J \sum_{\langle ij \rangle} n_i^d n_j^d 
  \: .
\label{eq:Heff}
\end{eqnarray}
Here, we have introduced the doublon creator
\be
d^{\dagger}_i = c^{\dagger}_{i\downarrow} c^{\dagger}_{i\uparrow} \: .
\ee
Furthermore, we have defined $n_i^d = d^\dagger_i d_i$ and $J=4T^2/U$. 
In the two-particle case, the interaction term is irrelevant, and the bound state has a tight-binding-like dispersion with effective doublon hopping $J$, see first term in Eq.\ (\ref{eq:Heff}).

Since we want to make contact with spectroscopic electron-emission experiments, such as valence band photoemission and CVV Auger electron spectroscopy, \cite{Pot01b} we consider single- and two-{\em hole} (opposed to electron) excitations in the following and restrict ourselves to the filling range $1\leq n \leq 2$. 
The previously discussed two-body limit is now given by two holes at $n=2$. ``Doublons'' will from now on refer to empty sites generated by application of $d_{i}$, so that the bound states in Fig.\ \ref{fig:doublonEigenvaluesBA} are given by localized two-hole wave functions separated from the two-hole continuum.

%---------------
\section{Decay products}
\label{sec:decay}
%---------------

Our main goal is to investigate the fate of a local doublon excitation of the form
\be
\ket{\Psi_i} = d_{i} \ket{0,N} \: ,
\ee
where $\ket{0,N}$ is the ground state in the subspace with given total particle number $N$ ($L\le N \le 2L$).
Alternatively, we may consider an excitation with a nonzero momentum, 
\be
\ket{\Psi\lr{k}} = d\lr{k} \ket{0,N} = \frac{1}{\sqrt{L}} \sum_i e^{ikR_i} d_{i} \ket{0,N} \: .
\ee
We start the discussion with the analysis of the decay products, i.e.\ the unbound eigenstates for the nontrivial case $n<2$, generalizing the results displayed in Fig.\ \ref{fig:doublonEigenvaluesBA}.

Formally, the initial state can be written as a linear combination of eigenstates.
If $\ket{m,N}$ denotes the $m$-th eigenstate of $H$ in the $N$-particle subspace, we have
\be
\ket{\Psi\lr{k}} = \sum_m \alpha_{m}^{\lr{N-2}}\lr{k} \ket{m,N-2} \: 
\ee
with some coefficients $\alpha_{m}^{\lr{N-2}}$.
As in the two-body example above, the excitations from the ground state $\ket{0,N}$, i.e.\ the final states $\ket{m,N-2}$, must be classified.
In particular, we need the information on whether the two added holes are bound or unbound. 
To this end, let us first define the accessible set $\mathcal{S}$ of unbound scattering states. 
Then the decay probability is found by projecting onto the space spanned by $\mathcal{S}$:
\be
p\lr{\Psi\lr{k}\rightarrow\mathcal{S}} 
= 
\sum_{m \in\mathcal{S}} \big|\alpha_{m}^{\lr{N-2}}\lr{k}\big|^2
\: .
\ee

%%%%%%%%%%%%%%%%%%%%%%%%%%%%%%%%%%%%%%%%%%%%%%%%%%
\begin{figure*}[t]
\includegraphics[width=0.9\textwidth]{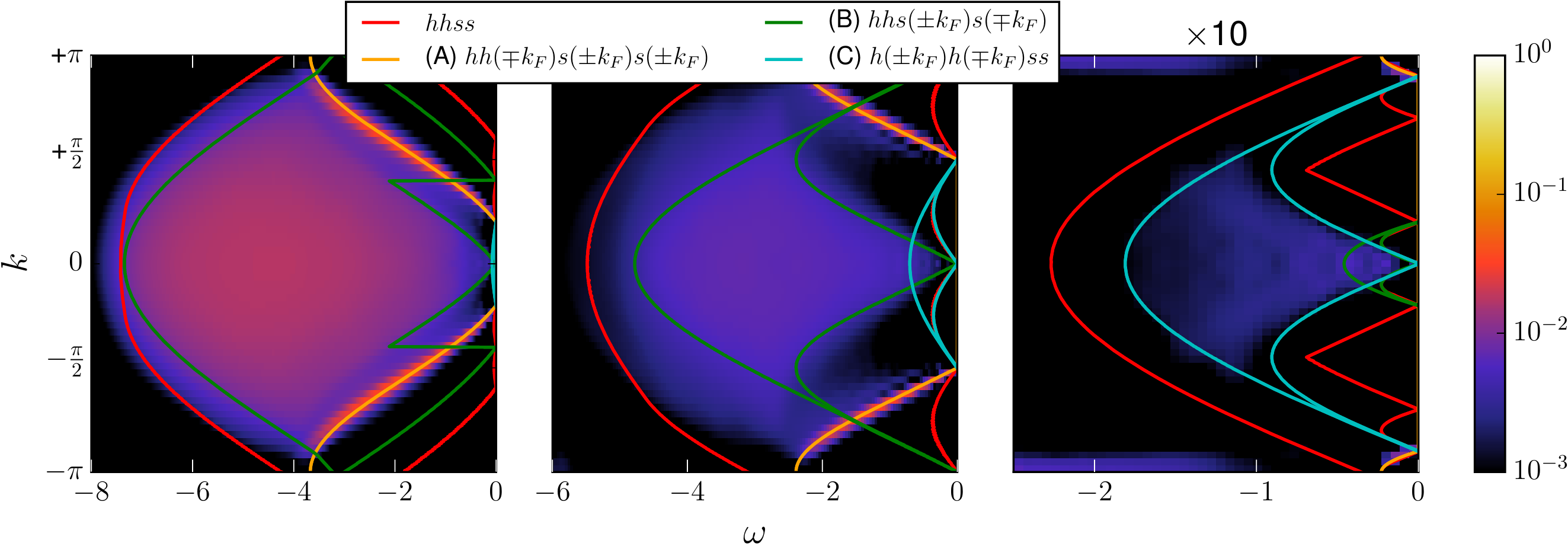}
\caption{
DMRG results (color code) for the band-like parts of the two-hole spectral function (\ref{eq:A}) for $L=60$ lattice sites, $U=6$ and the fillings $n=1.8$ (left), $n=1.5$ (middle), $n=1.1$ (right) with overlaid borders of the Bethe ansatz continua as coloured lines (see explanation in the text). The spectrum at $n=1.1$ has been multiplied by a factor of $10$ to maintain a common scale.
}
\label{fig:bandlikeBA}
\end{figure*}
%%%%%%%%%%%%%%%%%%%%%%%%%%%%%%%%%%%%%%%%%%%%%%%%%%

The Bethe ansatz \cite{Essler_2005} provides us with this classification: 
It turns out that all eigenstates can be constructed from elementary excitations of the (anti-)holon, the spinon, and a $k$-$\Lambda$ string of integer length. 
All of them have dressed momenta and energies and can be combined with each other to create particular physical excitations, see Eq.\ (\ref{eq:BA_p_eps}) below and Refs.\ \onlinecite{Essler_2005, Solyom_2010}.
For example, an electron hole created in a photoemission process consists of a holon and a spinon, while a spin flip in the dynamical spin structure factor (measured by neutron scattering) excites two spinons. 
A special case are the strings, which correspond to bound states of two or more electrons. 
In our case of a doublon excitation, we expect that $\mathcal{S}$ is given by a continuum of two holons and two spinons, with a more complicated kinematics than in the case of photoemission.

Unfortunately, the calculation of matrix elements within the Bethe ansatz is very difficult in practice (except for special limits).
It is therefore necessary to combine the Bethe ansatz with an essentially exact numerical approach such as the DMRG.
A quantity which is accessible to DMRG and has the relevant matrix elements is the two-hole spectral function
\begin{widetext}
\be
A_{\rm 2-hole}\lr{\omega,k} 
=
\frac1L
\sum_m 
\norm{
\matrixel{m,N-2}{\sum_{j} e^{ikj} c_{j\uparrow}c_{j\downarrow}}{0,N}
} 
\delta \lr{\omega + 2 \mu - \lr{E_0^{\lr{N}} - E_m^{\lr{N-2}} }}
\: .
\label{eq:A}
\ee
\end{widetext}
Evidently, it contains the coefficients $\big|\alpha_{m}^{\lr{N-2}}\lr{k}\big|^2$.
However, we now have the converse case that the character of the contributing eigenstates at a given $k$ and $\omega$ is unknown, so that a consultation of the Bethe ansatz is necessary to interpret the spectra.

%%%%%%%%%%%%%%%%%%%%%%%%%%%%%%%%%%%%%%%%%%%%%%%%%%
\begin{figure*}[t]
\includegraphics[width=0.75\textwidth]{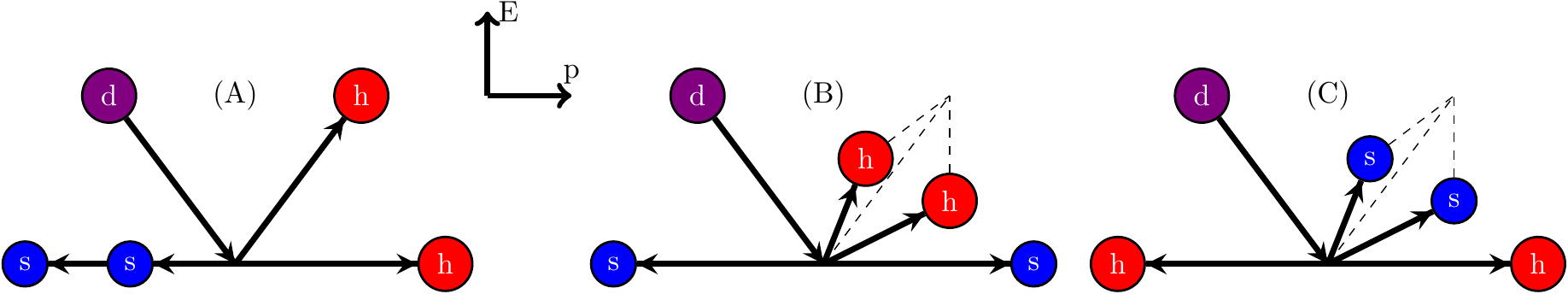}
\caption{
Sketches of the doublon decay processes, Eq.\ (\ref{eq:decayA}) and Eq.\ (\ref{eq:decayBC}), as determined from the band-like part in Fig.\ \ref{fig:bandlikeBA}. The horizontal axis shows the magnitude and direction of the involved momenta, the vertical axis the magnitude of the involved energies (such that a flat line corresponds to the Fermi energy). 
Purple circles: doublons. Red circles: holons. Blue circles: spinons.
}
\label{fig:decay}
\end{figure*}
%%%%%%%%%%%%%%%%%%%%%%%%%%%%%%%%%%%%%%%%%%%%%%%%%%

We have carried out a detailed analysis of $A_{\rm 2-hole}\lr{\omega,k}$ in a previous work,\cite{Rausch_Potthoff_2016} by using DMRG combined with the Chebyshev polynomial expansion. \cite{Silver_Roeder_1997, Weisse_etal_2006, Weisse_Fehske_2008} Essentially, two features can be observed: (i) a \emph{band-like part} with weak intensity coming from the scattering states and (ii) a \emph{satellite} with high spectral weight coming from the bound states. This dichotomy is more or less preserved for fillings $n<2$, although the spectra are complicated by two additional features: a doublon-hole continuum overlapping with the satellite and a quadruplon peak at higher binding energies for fillings $n\lesssim1.5$, representing a bound state of four holes. The bound quadruplon can be identified with the $k$-$\Lambda$ string of length 2 of the Bethe ansatz. At half filling, all the mentioned features coincide in a single peak (see Ref.\ \onlinecite{Rausch_Potthoff_2016} for details).

Here, we look more closely at the band-like part, which is expected to correspond to the set of unbound scattering states $\mathcal{S}$ introduced above.
To verify this assumption, we plot the intensity given by Eq.\ (\ref{eq:A}), as obtained from the DMRG, in the $\omega$-$k$ plane and compare with the momenta and excitation energies from the Bethe ansatz. 
The latter are given by:
\begin{widetext}
\be
\begin{split}
p_{\text{tot}}\lr{k_1,k_2,k_3,k_4} &= p^h_1 + p^h_2 + p^s_1 + p^s_2 = p^h\lr{k^h_1} + p^h\lr{k^h_2} + p^s\lr{k^s_1} + p^s\lr{k^s_2}, \\
\epsilon_{\text{tot}}\lr{p_{\text{tot}}} &= \epsilon^h\lr{p^h_1} + \epsilon^h\lr{p^h_2} + \epsilon^s\lr{p^s_1} + \epsilon^s\lr{p^s_2} 
\: .
\end{split}
\label{eq:BA_p_eps}
\ee
\end{widetext}
Here, $k^h_i$ and $k^s_i$ are the holon and spinon wave vectors, respectively; $p^h\lr{k^h}$ and $p^s\lr{k^s}$ are the corresponding dressed momenta, and $\epsilon^h\lr{p^h}$ and $\epsilon^s\lr{p^s}$ are the corresponding dressed energies. When all the wavevectors are varied in their respective ranges, one obtains the 2-holon-2-spinon continuum (we use the shorthand $hhss$), which is expected to delimit the whole spectral support of the band-like part. However, substructures are possible when some of the momenta are fixed at certain values. 
If three of the four momenta are fixed, a dispersion relation rather than a continuum is obtained.

Fig.\ \ref{fig:bandlikeBA} displays the band-like part as obtained from the DMRG for fillings $n=1.8$, $n=1.5$ and $n=1.1$ (see the color code). 
The calculations have been performed at $U=6$ for a system with $L=60$ lattice sites using open boundary conditions. 
The red lines, as obtained from the BA, indicate the boundaries of the $hhss$ continuum.

We observe that they indeed enclose the band-like part completely, except for some spillage of spectral weight at $n=1.8$ close to $k=0$ and $\omega=-8T$ due to numerical broadening and finite-size effects. 
Notably, there are also regions within the $hhss$ continuum without any spectral weight at all, particularly close to half filling.

First of all, we find an intense ridge-like structure beyond which the spectral weight suddenly drops off.
This is best seen for $n=1.8$ where it starts at $k = \pm 0.2 \pi$ and $\omega=0$ and disperses to the zone boundary.
With decreasing filling, its starting point shifts closer and closer toward the zone edge, and for $n=1.1$ it only shows up faintly as two speckles of spectral weight at about $k=\pm0.9\pi$. 
Using the BA, this structure can be identified as the decay channel
\be
\text{(A)} \qquad d \to h\lr{k} ~ h(\pm k_F^h) ~ s(\mp k_F^s) ~ s(\mp k_F^s) \: ,
\label{eq:decayA}
\ee
namely as a final state at filling $n$ where a single holon is fixed to its Fermi momentum $k_F^h=n\pi\mod\pi$, and where both of the spinons are also fixed to their respective Fermi momenta $k_F^s = \frac{n\pi}{2}\mod\pi$, but with signs opposite to $k_F^h$. 
Since we have that $k_F^s=k_F^h/2$ for any filling, these three momenta cancel each other, so that one is effectively left with a single dispersing holon $h(k)$. 
This is indicated by the orange line in the plot.

The other two relevant decay channels are found to be
\be
\begin{split}
\text{(B)} \qquad d &\to h(k_1) ~ h(k_2) ~ s(\pm k_F^s) ~ s(\mp k_F^s) \: , \\
\text{(C)} \qquad d &\to h(\pm k_F^h) ~ h(\mp k_F^h) ~ s(k_1) ~ s(k_2),
\label{eq:decayBC}
\end{split}
\ee
namely, final states where either the two holons or the two spinons are pinned to their Fermi momenta with opposite signs. In these cases, the fixed momenta also cancel each other, while the remaining two dispersive particles form a continuum. In the case of (B), all the energy is carried by the holons, in the case of (C) by the spinons. The respective continuum boundaries are indicated by the cyan and green boundary lines in Fig.\ \ref{fig:bandlikeBA}. A visualization of the three decay channels is provided in Fig.\ \ref{fig:decay}.

Since spinons only start to become relevant toward half filling, the area (or phase space) of the decay channel (C) is vanishingly small at $n=1.8$. 
It becomes noticeable at $n=1.5$, but does not seem to coincide with a specific part of the DMRG spectrum. 
At $n=1.1$, however, its shape coincides with the whole support of the band-like part apart from the speckles of channel (A) at $k=\pm0.9\pi$ mentioned above. 
We thus conclude that (C) becomes the dominant decay channel close to half filling. 

On the other hand, holon excitations disappear toward half filling, so that the phase space related to (B) becomes rather small for $n=1.1$ while it is much more extended at $n=1.8$ and $n=1.5$.
One observes that its lower border reproduces the onset of the spectral weight close to the $\Gamma$-point rather well, while its upper border does not seem to delimit anything specific.

We conclude that the picture of a doublon excitation decaying with a certain probability into unbound scattering states remains valid at intermediate fillings. 
This process shows up as the band-like part of the AES two-hole spectrum. 
However, in stark contrast to the physics of the simple $n=2$ limit, the analysis of the substructure of this continuum clearly reveals that the decay products must be interpreted in terms of spinons and holons rather than electron holes in case of a one-dimensional lattice.
There are no indications that we can describe the decay in terms stable (or metastable) unbound electron holes.

%---------------
\section{Real-time dynamics of the decay}
\label{sec:dynamics}
%---------------

The continuum of scattering states results from the dispersion of the unbound excitations that are left after the doublon decay. 
This decay, however, is expected to be incomplete, since the kinematic constraints resulting from energy and momentum conservation will block low-order scattering processes on an intermediate time scale, while high-order processes only become relevant for very long times scales. 
Doublon decay is possible via low-order processes at extremely short times $\tau$. 
Assuming that the variance of the total energy in the initial state scales linearly with $U$ (for strong $U$), the Mandelstam-Tamm energy-time uncertainty relation \cite{MS45} predicts $\tau \sim h/U$.
This expectation can be checked by means of systematic DMRG calculations using real-time propagation in the entire filling range ($1\leq n \leq 2$).

%%%%%%%%%%%%%%%%%%%%%%%%%%%%%%%%%%%%%%%%%%%%%%%%%%%%%%%%
\begin{figure}[t]
\includegraphics[width=0.95\columnwidth]{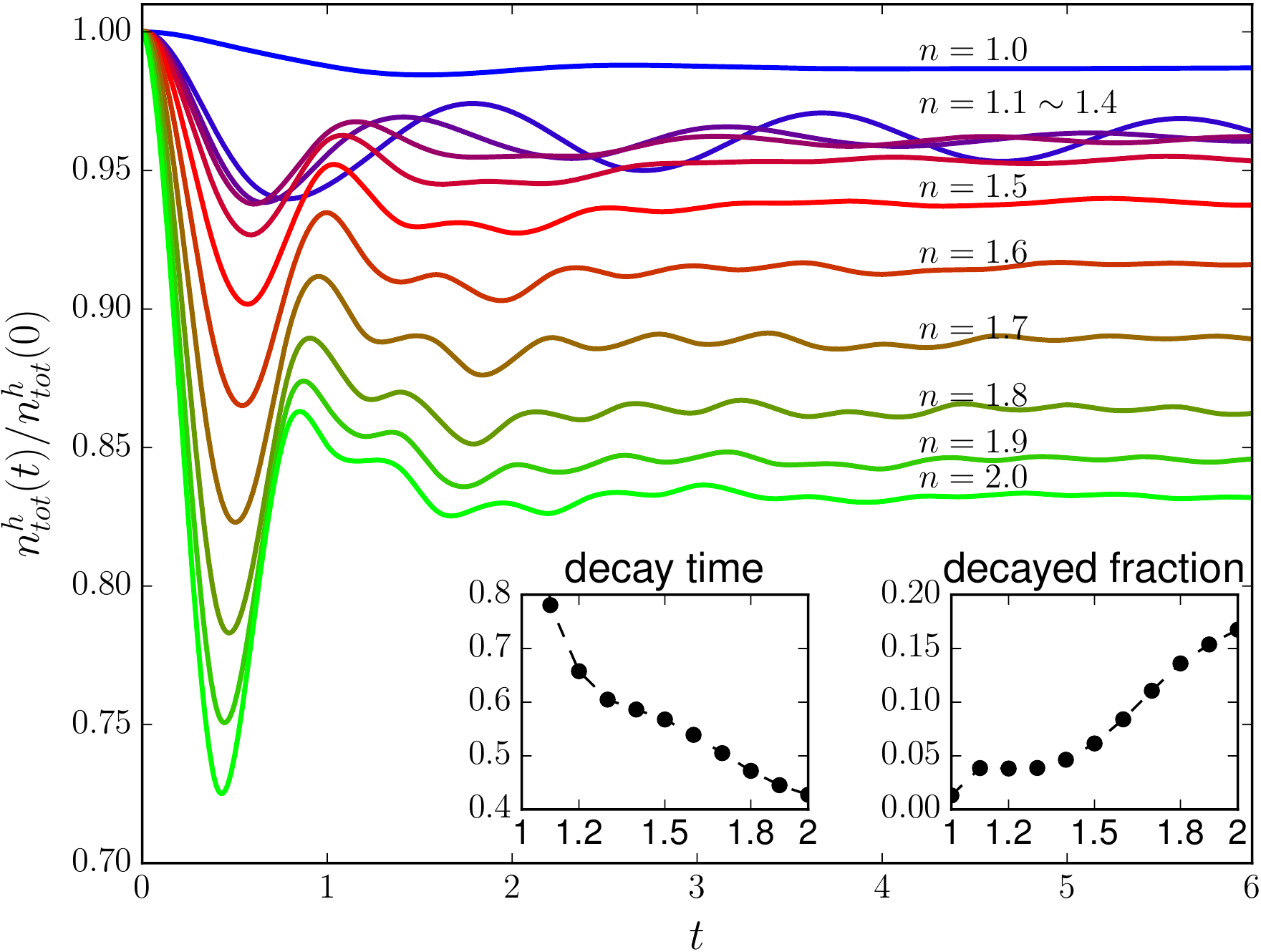}
\caption{Total hole density $n^{h}_{\text{tot}} (t)$, normalized to $n^{h}_{\text{tot}} (0)$, for various fillings as indicated. Time-dependent DMRG calculation at $U=6$ for $L=40$. 
Left inset: characteristic decay time $t_{\rm min}$, defined as the location of the first minimum, as a function of filling. 
Right inset: decayed fraction of doublons, defined as the time average $1 - \overline{n^{h}_{\text{tot}} (t) / n^{h}_{\text{tot}} (0)}$ between $t=4$ and the maximal propagation time, as a function of filling.
}
\label{fig:htot}
\end{figure}
%%%%%%%%%%%%%%%%%%%%%%%%%%%%%%%%%%%%%%%%%%%%%%%%%%%%%%%%

As discussed in the introduction, there are several ways to initiate the process. 
Here, we study the dynamics following a sudden doublon creation at a site $i_{0}$, i.e.\ we start from a normalized state $\ket{\Psi_{i_0}} = (1/\sqrt{\mathcal{N}}) c_{i_0\uparrow}c_{i_0\downarrow} \ket{0,N}$. 
Compared to a geometrical quench, this has the disadvantage of having somewhat less control over the initial occupancies, since the doublon excitation will be slightly spread over several lattice sites. 
The advantage is, however, that the initial state is the same as in Eq.\ (\ref{eq:A}), and thus its propagation in real time offers a complementary view on the same problem.
The time evolution of the initial state $\ket{\Psi\lr{t}} = e^{-iHt} \ket{\Psi_{i_0}}$ is computed using the DMRG combined with the two-site variant of a time-evolution algorithm based on the time-dependent variational principle.\cite{Haegeman_etal_2014} 

One clearly has to distinguish between doublon decay and doublon propagation. 
For the latter, it is important to glean spatially resolved information from the wave function. 
This can be done by calculating expectation values $\avg{O_j}\lr{t} = \matrixel{\Psi\lr{t}}{O_j}{\Psi\lr{t}}$ of suitable local observables $O_j$.
Obvious choices for $O_{j}$ include the charge density $n_j = n_{j\uparrow} + n_{j\downarrow}$ or observables resulting from the decomposition $1 = n_j^d + n_j^h + n_j^s$, which allows us to specifically look at doubly occupied sites $n^d_j = n_{j\uparrow}n_{j\downarrow}$, empty sites $n^h_j = n^d_j-n_j+1$ or singly occupied sites $n^s_j=1-n^d_j-n^h_j = n_j-2n_{j\uparrow}n_{j\downarrow}$. 
For the given problem, $n^h_j$ is most useful, indicating both the decay and the propagation of the doublon excitation.
Note that while the expression for $\avg{n^h_j}\lr{t}$ looks similar to the real-time representation of the two-hole spectral function (\ref{eq:A}), there is no simple transformation between the two (except for the full and empty band),\cite{Hofmann_Potthoff_2012} so that the real-time dynamics provides us with additional information extracted from the same wavefunction.

%%%%%%%%%%%%%%%%%%%%%%%%%%%%%%%%%%%%%%%%%%%%%%%%%%%%%%%%
\begin{figure}[t]
\includegraphics[width=0.95\columnwidth]{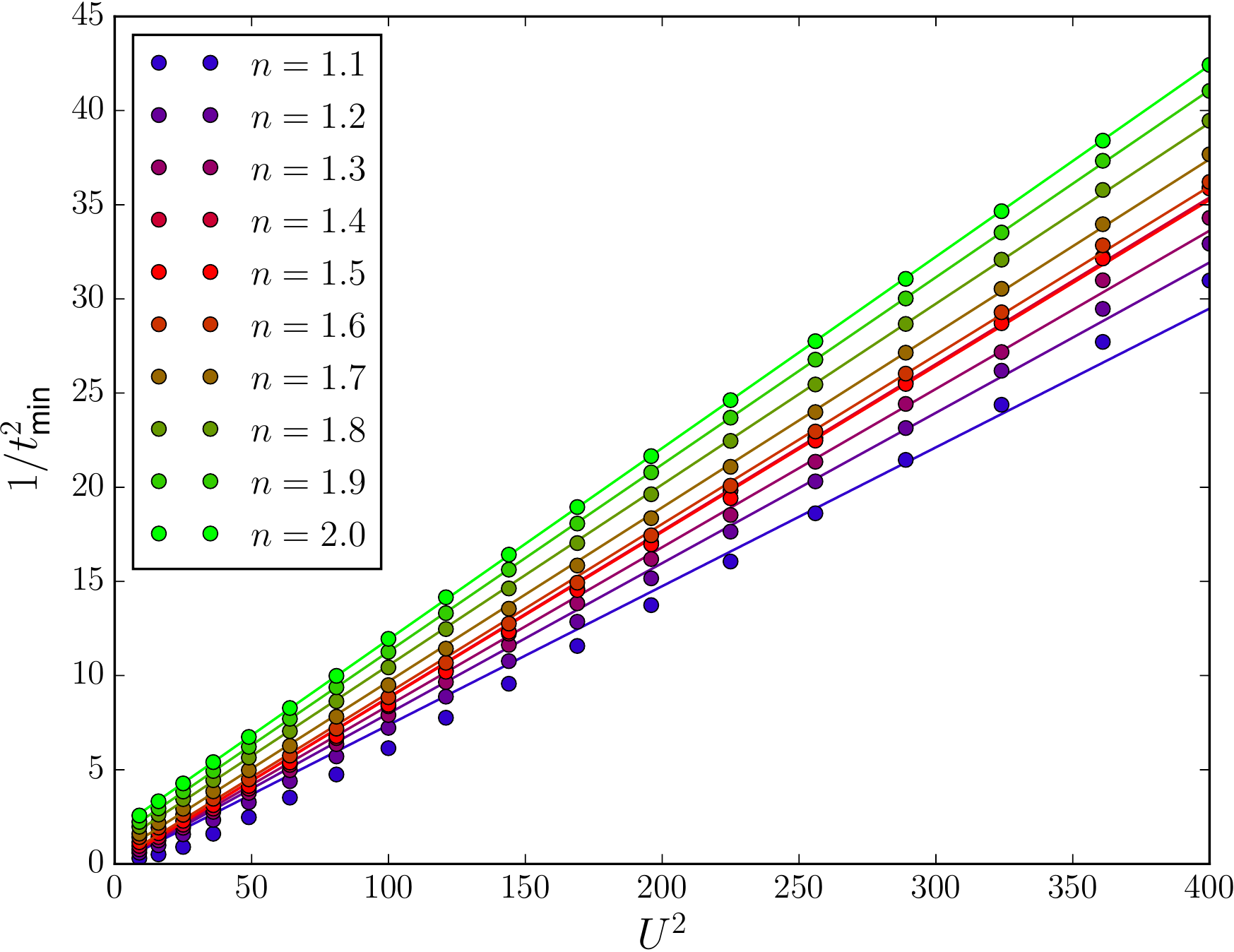}
\caption{
Inverse square of the decay time $t_{\rm min}$, where $n^{h}_{\text{tot}}(t)$ exhibits its first pronounced minimum, as a function of $U^{2}$ for different fillings.
Lines: linear fits of the DMRG data based on the result of the two-site model, Eq.\ (\ref{eq:omega}), see text.
}
\label{fig:dip}
\end{figure}
%%%%%%%%%%%%%%%%%%%%%%%%%%%%%%%%%%%%%%%%%%%%%%%%%%%%%%%%

Let us first concentrate on the doublon decay and discuss the {\em total} hole density 
\be
n^{h}_{\text{tot}} (t) = \sum_{j} \avg{{n}^h_j} (t)
\ee
in the state $\ket{\Psi(t)}$, as obtained from time-dependent DMRG. 
Fig.\ \ref{fig:htot} shows $n^{h}_{\text{tot}} (t)$ (normalized to $n^{h}_{\text{tot}} (0)$) for different fillings. 
We find a sharp dip for very short times and a subsequent ``relaxation'' to an intermediate value with weak superimposed oscillations, very similar to the overall behavior that has been seen for the two-particle case (here corresponding to $n=2$) in Ref.\ \onlinecite{Hofmann_Potthoff_2012}. 
The position of the first minimum $t_{\rm min}$ can be interpreted as the characteristic ``decay time.'' 
The drop to the emerging constant value indicates the decayed fraction of doublons $r_{\rm decay}$. 

Similarly to the two-particle case, the decay time $t_{\rm min}$ is rather short, below one inverse hopping at $U=6$, and only slightly increases with decreasing filling.
The $U$-dependence of the decay time is also instructive and is shown in Fig.\ \ref{fig:dip}.
For $n=2$, an effective two-site model can be employed, since on the very short decay time scale, only the neighboring site can be explored by the doublon. 
Furthermore, the initial state of a two-site model is actually locally the same as in the $n=2$ case, consisting out of an empty site with an adjacent doubly occupied one in the former case, and a whole doubly occupied lattice in the latter case.
The calculation for the two-site model predicts a collapse-and-revival frequency $\omega_{0} = \sqrt{U^2+16T^2}$. \cite{Kajala_etal_2011} 
Indeed, the data for the first minimum at $n=2$, which we expect to be at $\omega_{0}t_{\rm min}=\pi$ for a cosine, are nicely fitted by 
\be
1/t_{\rm min}^{2} \approx \lr{U^{2} + 16 T^{2}}/\pi^2 \: .
\label{eq:omega}
\ee 
Obviously, the simpler proportionality $t_{\rm min}\sim 1/U$ emerges for $U \gg T$.
For $n<2$ down to $n \approx 1.4$, we find that there is still a $U^2$ dependence of $1/t_{\rm min}^{2}$ (shown by the linear fits in Fig.\ \ref{fig:dip}), reflecting the fact that the doublon decay is more and more suppressed with increasing $U$.
Note, however, that discrepancies with respect to the fit grow in the range $1 < n \lesssim 1.4$.

On the intermediate time scale up to the maximum propagation time the main effect of the decreasing filling seems to consist in the decrease of the decayed fraction of doublons $r_{\rm decay}$, see the left inset in Fig.\ \ref{fig:htot}.
This indicates that the doublon is actually stabilized by the presence of more and more singly occupied sites.
Indeed, there is an intuitive argument explaining why the phase space for doublon decay should shrink: 
With decreasing filling, the presence of singly rather than doubly occupied sites next to the doublon becomes more and more likely for typical configurations contributing to $c_{i_0\uparrow}c_{i_0\downarrow} \ket{0,N}$. 
However, configurations with a neighboring doubly occupied site are required to fill the empty site without creating another empty one.
This argument also consistently explains the dramatic overall loss of spectral weight in the band-like part of the two-hole spectral function with decreasing $n$ (see Fig.\ \ref{fig:bandlikeBA}).

%%%%%%%%%%%%%%%%%%%%%%%%%%%%%%%%%%%%%%%%%%%%%%%%%%%%%%%
\begin{figure}[t]
\includegraphics[width=0.95\columnwidth]{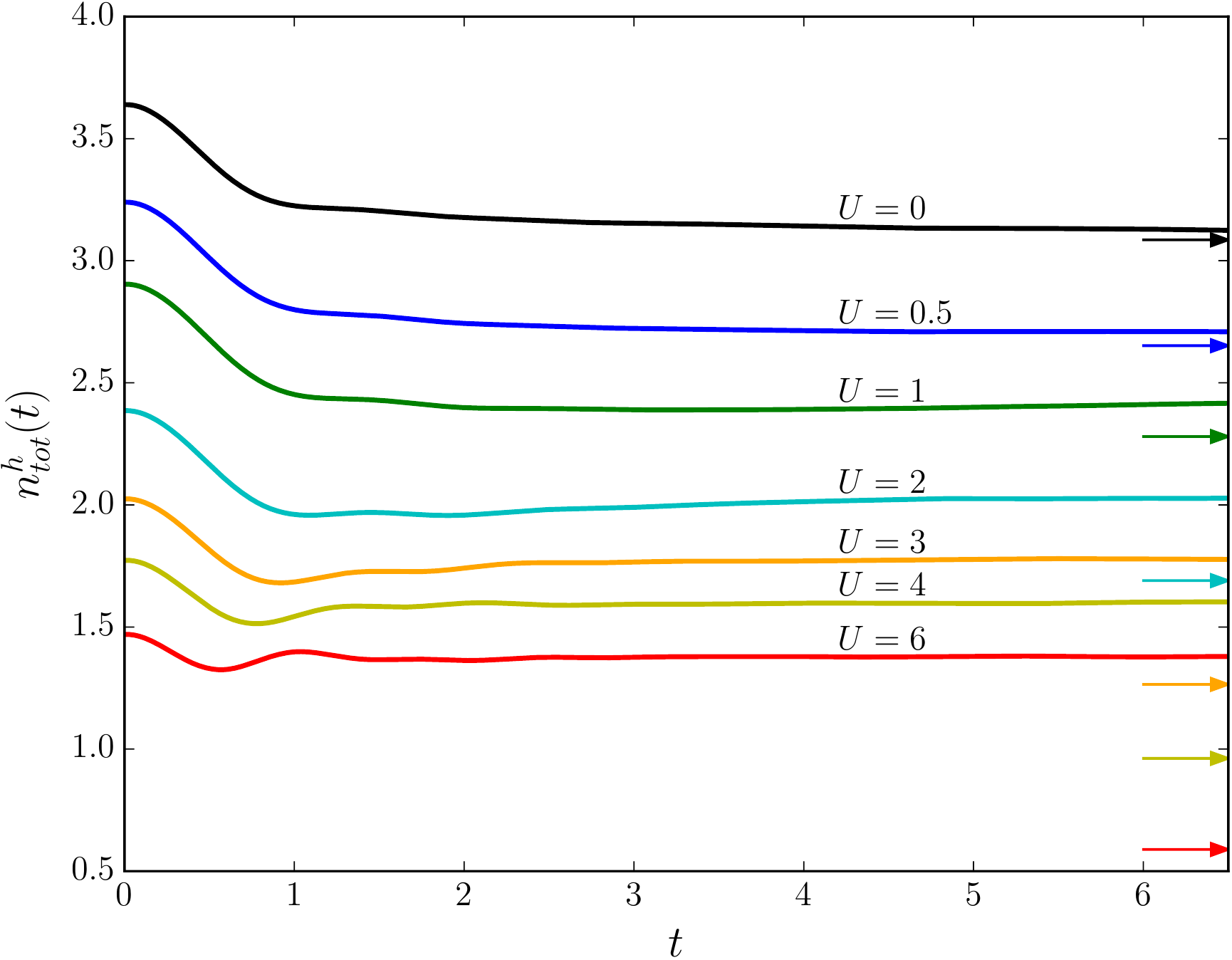}
\caption{
Comparison of the unnormalized hole density $n^{h}_{\text{tot}} (t)$ with the expected relaxed ground state expectation value (indicated by arrows of the same colour) for various values of $U$ at a filling of $n=1.5$ and $L=40$.
}
\label{fig:hrelax_L=40_n=1.5}
\end{figure}
%%%%%%%%%%%%%%%%%%%%%%%%%%%%%%%%%%%%%%%%%%%%%%%%%%%%%%%

At half filling the doublon is most stable. 
At the same time, the {\em creation} of a doublon is most unlikely in this case:
This is because for $n=1$, we have $c_{i_0\uparrow}c_{i_0\downarrow} \ket{0,N} \ne 0$ only for ground-state configurations with a doubly occupied site $i_{0}$. 
This requires a virtual hopping process involving an energy of the order of $U$, which, for strong $U$, is strongly suppressed.
In the two-hole spectral function this results in the overall decrease of total spectral weight with filling (including the doublon satellite). \cite{Rausch_Potthoff_2016}
Note that this effect cannot be seen in Fig.\ \ref{fig:htot}, since it is compensated by the normalization constant $n^{h}_{\text{tot}}(0)$, which also becomes small at (and close to) $n=1$. 

After the initial decay on the time scale $t_{\rm min}$, one observes oscillations of $n_{\rm tot}^{h}(t)$ in Fig.\ \ref{fig:htot} resulting from repeated decay-and-recombination processes.
The corresponding frequency is of the order of $U$.
These oscillations die out and the decay process is essentially completed within a few inverse hoppings.
However, the filling regime $1 < n \lesssim 1.4$ is again somewhat exceptional. 
Here, the decayed fraction is seen to reach a plateau at a value of about $0.05$ (see right inset). 
Furthermore, the dip of $n_{\rm tot}^{h}(t)$ at $t_{\rm min}$ is so shallow that the ``decay time'' defined via its position does not seem to be very meaningful anymore. 
Instead, $n^{h}_{\text{tot}} (t)$ shows several oscillations of comparable magnitude and the decay process does yet not seem to be completed.

We conclude that our data confirm the physical picture of the doublon decay outlined above:
The more or less constant value of $n^{h}_{\text{tot}} (t)$ on the intermediate time scale is the result of the kinematic constraints becoming active, while an initial decay is possible on the short time scale $t_{\rm min} \sim 1/U$ via low-order scattering processes.
Note that the ``relaxed'' values of $n_{\rm tot}^{h}(t)$ that can be read off as time averages from the DMRG data for the different fillings in Fig.\ \ref{fig:htot} do not coincide with the respective ground-state expectation values $\matrixel{0,N-2}{n^h}{0,N-2}$ in the subspace with $N-2$ electrons as has been checked numerically.
The latter is expected to be reached for $t \to \infty$, probably on a time scale exponentially long in $U$, on which higher-order scattering processes are activated.
However, this time scale is inaccessible to DMRG.

One can check that the ``relaxed'' value of $n_{\rm tot}^{h}$ more and more approaches the ground-state expectation value $\matrixel{0,N-2}{n^h_{\text{tot}}}{0,N-2}$ when $U$ is decreased.
This is shown in Fig.\ \ref{fig:hrelax_L=40_n=1.5} for $n=1.5$.
It is noticeable that even small values of $U$ have the consequence that the relaxed value is not completely reached within the accessible time scale.
The hole density actually seems to go \emph{away} from the relaxed value for some $U$, which is probably the result of an overlaid long-time oscillation, possibly with a frequency of the order of $J$.
Generally, the anticipated relaxation of expectation values of local observables to their ground-state values in a setup with a locally perturbed initial ground state is conceptually similar to the eigenstate thermalization hypothesis, \cite{Deu91,Sre94,RDO06} but somewhat simpler. 
Still, this requires further investigations addressing, for example, the role of the Bethe-ansatz integrability of the model. 

%---------------
\section{Doublon propagation}
\label{sec:prop}
%---------------

\begin{figure}[t]
\includegraphics[width=0.95\columnwidth]{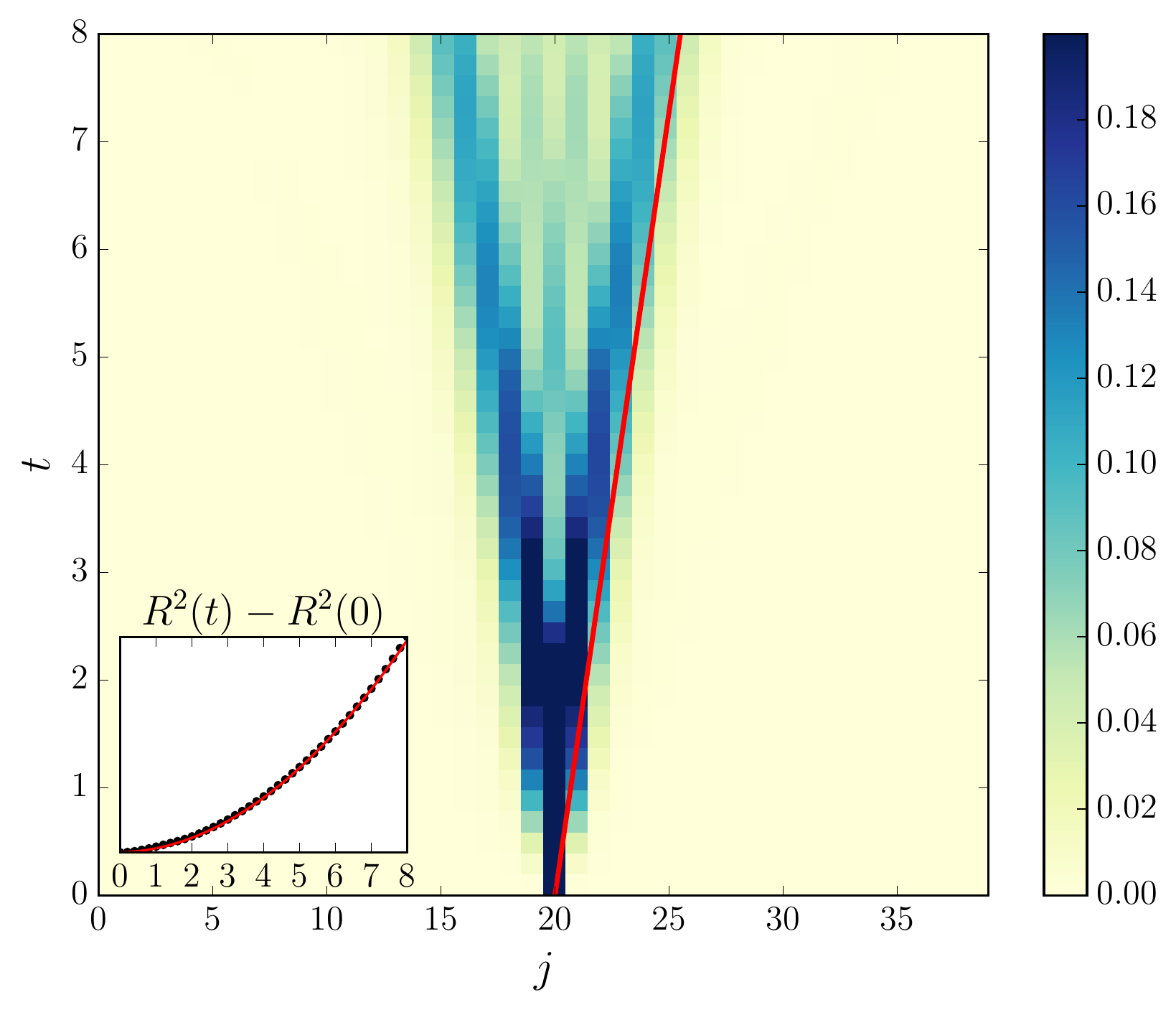}
\caption{
Space-time plot of $\avg{\tilde{n}^h_j}\lr{t}$ $n=1.8$. 
DMRG calculation at $U=6$ for $L=40$.
The inset shows $R^2\lr{t}$ (after subtracting the initial value) with the fit according to Eq.\  (\ref{eq:Rfit}). 
The red line in the main plot indicates $vt$ with the velocity extracted from the fit.
}
\label{fig:h_n=1.8}
\end{figure}

\begin{figure}[t]
\includegraphics[width=0.95\columnwidth]{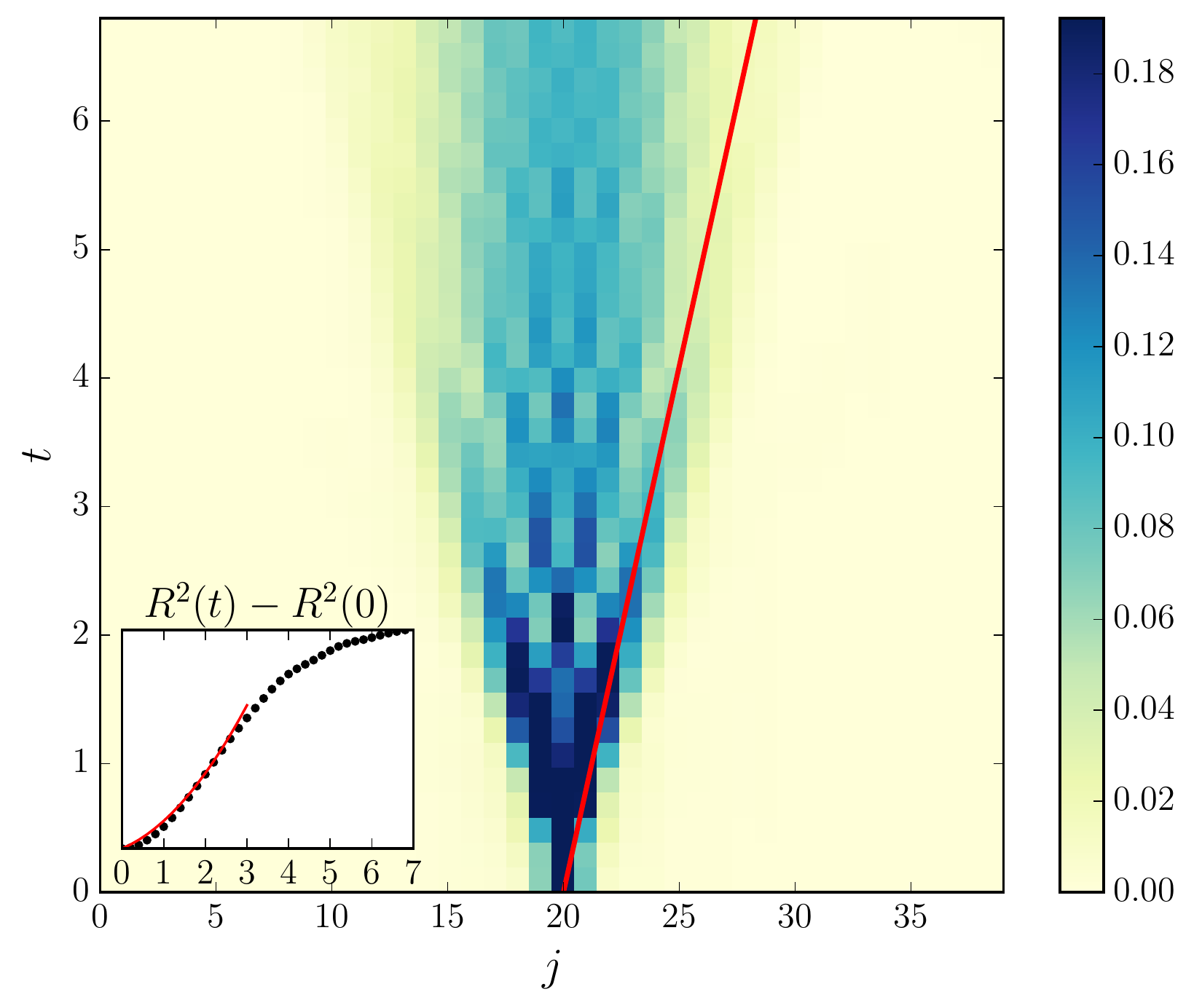}
\caption{
Space-time plot of $\avg{\tilde{n}^h_j}\lr{t}$ for $n=1.2$ and other parameters as in Fig.\ \ref{fig:h_n=1.8}. 
The fit is limited to $t=3$.
}
\label{fig:h_n=1.2}
\end{figure}

Once created, the doublon tends to delocalize.
Its spatiotemporal evolution adds further complementary pieces of information to the physical picture. 
The result of a corresponding time-dependent DMRG calculation is shown in Fig.\ \ref{fig:h_n=1.8} for a filling of $n=1.8$ and in Fig.\ \ref{fig:h_n=1.2} for $n=1.2$. 

In order to analyze the emerging light-cone dynamics of the initially localized excitation, two models can be employed: a ballistic and a diffusive one. 
Broadly speaking, the former is a signature of a freely propagating excitation as in a noninteracting system, while scattering induced by interactions is essential for the latter. 
Although integrable one-dimensional systems with many conserved quantities, such as the Hubbard chain, are expected to behave ballistically as noninteracting ones, ballistic and diffusive transport have also been shown to coexist. \cite{Sirker_etal_2009, Andraschko_Sirker_2015}
We follow preceding work \cite{Dunlap_Kenkre_1986, Langer_etal_2009, Vidmar_etal_2013, Andraschko_Sirker_2015} in establishing the two models. 
For a given arbitrary localized density distribution $\avg{\rho_j}\lr{t}$ at time $t$ around the center of excitation $i_0$, we introduce the radius $R(t)$ via 
\be
R^2\lr{t}\big[\avg{\rho}\big] = \sum_j \lr{j-i_0}^2 \avg{\rho_j}\lr{t} \: .
\label{eq:Rsq}
\ee
With this we can distinguish between ballistic and diffusive cases.

%%%%%%%%%%%%%%%%%%%%%%%%%%%%%%%%%%%%%%%%%%%%%%%%%%%%%%%
%%%%%%%%%%%%%%%%%%%%%%%%%%%%%%%%%%%%%%%%%%%%%%%%%%%%%%%

Let us first consider the ballistic model. 
In general, for noninteracting particles with dispersion $\epsilon\lr{k} = \pm v\cos\lr{k} $ and an infinite lattice, a straightforward calculation yields that the corresponding expectation value of the local density is given by
\be
\avg{\rho_j}\lr{t} = \mathcal{J}_{j-i_0}^2\lr{vt} \: ,
\ee
where $\mathcal{J}_{n}\lr{x}$ is the $n$-th Bessel function of the first kind. 
Plugging this into (\ref{eq:Rsq}), we obtain
\be
R^2\lr{t}\big[\avg{\rho}\big] = \frac{1}{2} v^2t^2 \: .
\ee

In our case, we consider the doublon density $\avg{n^h_j}\lr{t}$, and the above relation should be fulfilled for large $U$ in the two limiting cases: 
On the one side we have $n=2$, where the effective doublon model (\ref{eq:Heff}) describes a free particle with $\epsilon\lr{k} = J\cos\lr{k}$. On the other side, we have $n=1$, where the effective model is the $t$-$J$ model \cite{CSO78} with vanishing $J$, so that an empty site corresponds to a hole in the singly-occupied antiferromagnetic background and similarly moves like a free particle with $\epsilon\lr{k} = -2T \cos\lr{k}$.

%%%%%%%%%%%%%%%%%%%%%%%%%%%%%%%%%%%%%%%%%%%%%%%%%%%%%%%
%%%%%%%%%%%%%%%%%%%%%%%%%%%%%%%%%%%%%%%%%%%%%%%%%%%%%%%

We now turn to the diffusive model. \cite{Langer_etal_2009, Vidmar_etal_2013, Andraschko_Sirker_2015} A solution of the continuous diffusion equation
\be
\pd{\rho}{t} - \mathcal{D} \pd{^2\rho}{x^2} = 0 \: , 
\ee
where $\mathcal{D}$ is the diffusion constant, with the boundary condition $\rho\lr{x,t=0} = \delta\lr{x}$ and with normalization $\int_{-\infty}^{\infty} dx~ \rho\lr{x,t} = 1$ is given by: 
\be
\rho\lr{x,t} = \rec{\sqrt{2\pi\mathcal{D}t}} ~ e^{-\frac{x^2}{4\mathcal{D}t}} \: .
\label{eq:rho_xt}
\ee
In this case, $R^2\lr{t}$ is just the second moment of the Gaussian distribution (\ref{eq:rho_xt}), given by the square of the standard deviation
\be
R^2\lr{t}\big[\rho\big] = 2\mathcal{D}t \: ,
\ee
so that the spread is slower and scales with $t$ rather than with $t^2$. 

For our problem, we thus make the ansatz
\be
R^2\lr{t}\big[\avg{n^h}\big] = \frac{1}{2} v^2t^2 + 2\mathcal{D}t
\label{eq:Rfit}
\ee
and fit the calculated $R^2\lr{t}\big[\avg{n^h}\big]$ to obtain the corresponding velocity $v$ and diffusion constant $\mathcal{D}$. 
To neutralize Friedel oscillations originating from open boundary conditions in the DMRG calculation, we subtract the ground-state expectation value and from now on work with 
\be
\avg{\tilde{n}^h_j}\lr{t} \equiv \matrixel{\Psi\lr{t}}{n^h_j}{\Psi\lr{t}}-\matrixel{N,0}{n^h_j}{N,0} \: .
\ee
These fits are shown in the insets of Figs.\ \ref{fig:h_n=1.8} and \ref{fig:h_n=1.2}, and the corresponding wavefronts $j\lr{t}=i_{0}+vt$ with $v$ determined by the fit, are indicated by red lines in the space-time plots. 
It turns out that at $n=1.2$, the fit only gives meaningful values up to a time of $t\sim3$; thereafter the dynamics changes noticeably. 
We therefore limit the fit to $t \le 3$ in this case and will discuss this effect further below.

\begin{figure}[t]
\includegraphics[width=0.95\columnwidth]{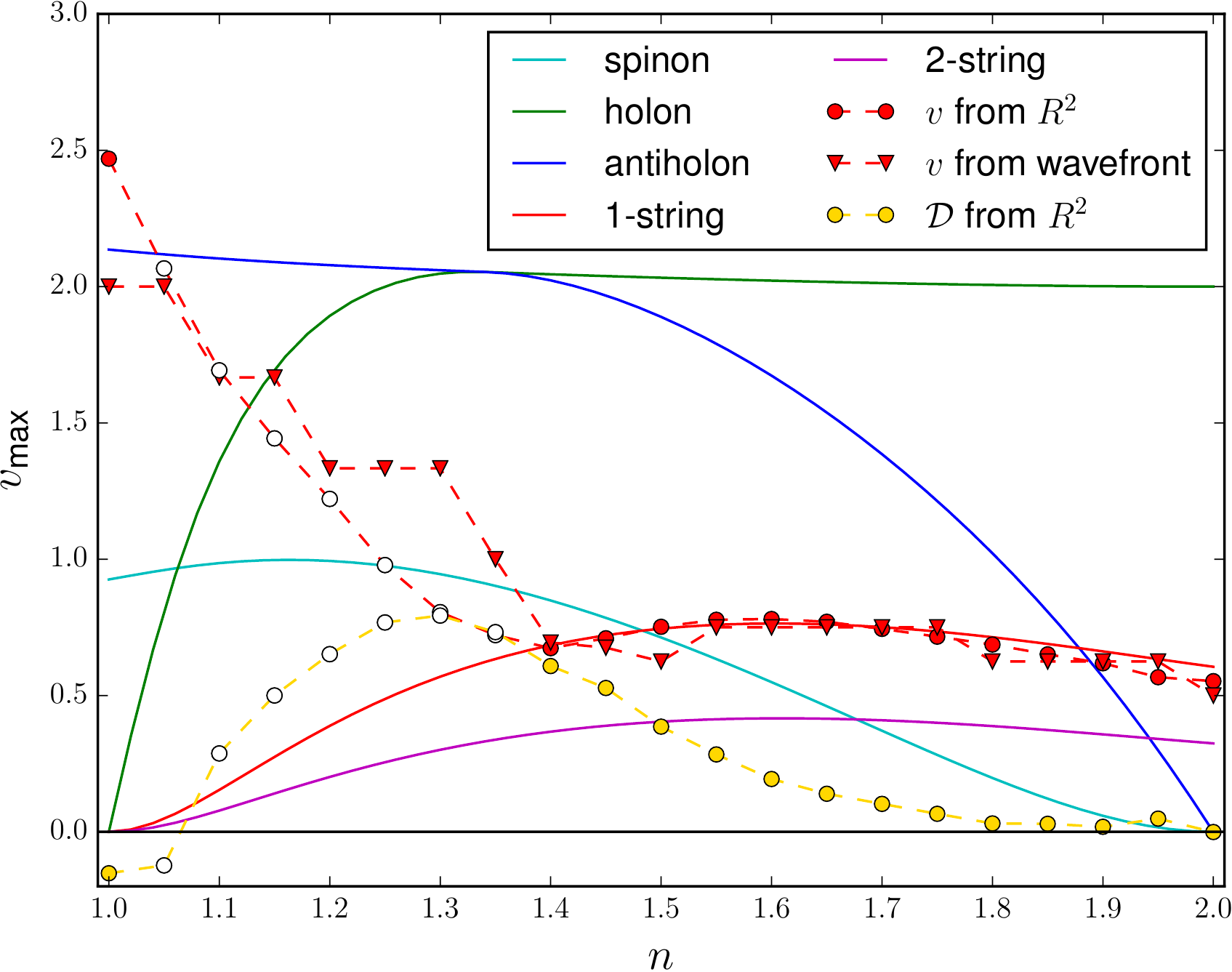}
\caption{
{\em Symbols:} 
velocity $v$ of the holon wavefront extracted from the fit of $R^2\lr{t}$ via Eq.\ (\ref{eq:Rfit}) (red circles) and directly from the DMRG data (red triangles). 
Yellow circles show the diffusion constant $\mathcal{D}$ from the same fit.
White circles indicate where the fitting procedure for $v$ and $\mathcal{D}$ was restricted to times $t\leq3$ (see text). 
{\em Lines:}
maximal velocities obtained via Eq.\ (\ref{eq:vmax}) from different Bethe ansatz eigenstates as functions of the filling.
}
\label{fig:velo}
\end{figure}

Fig.\ \ref{fig:velo} displays the filling dependence of the velocity $v$ (red circles) and of the diffusion constant 
$\mathcal{D}$ (yellow circles) as obtained from the fit (\ref{eq:Rfit}). 
The regime where Eq.\ (\ref{eq:Rfit}) is only valid up to $t\sim3$ is found to lie between $n=1$ and $n\sim1.4$, and the corresponding restricted fit results for $v$ and $\mathcal{D}$ are displayed as white circles.

For an interpretation of the obtained velocities, we once more make a comparison with the predictions of the Bethe ansatz.
To this end we write the time-dependent expectation value of $n^h_j$ in the following form using Bethe-ansatz eigenstates $\ket{k_n}$: \cite{Ganahl_etal_2012}
\begin{widetext}
\be
\begin{split}
\avg{n^h_j}\lr{t} &= \rec{\mathcal{N}} \sum_{p_mk_n} \matrixel{0}{d_{i_0}^\dagger}{p_m} \matrixel{k_n}{d_{i_0}}{0} \matrixel{p_m}{e^{ip{i_0}} e^{iHt} e^{-iP\lr{j-{i_0}}} O_{i_0} e^{+iP\lr{j-{i_0}}} e^{-iHt} e^{-ik{i_0}}}{k_n}  \\
                  &= \rec{\mathcal{N}} \sum_{p_mk_n} \matrixel{0}{d_{i_0}^\dagger}{p_m} \matrixel{p_m}{O_{i_0}}{k_n} \matrixel{k_n}{d_{i_0}}{0} e^{i\phi_{mn}\lr{p,k}},
\end{split}
\ee
\end{widetext}
where $P$ is the momentum operator, and $\phi_{mn}\lr{p,k}$ is a plane-wave-like phase:
\be
\phi_{mn}\lr{p,k} = \lr{k-p} \lr{j-{i_0}} - \lr{E_n \lr{k} - E_m \lr{p}} t \: .
\ee
A stationary phase requires $\partial \phi_{mn}\lr{p,k} / \partial k = 0$. 
This implies that the wave with wave vector $k$ travels the distance $\big|j-{i_0}\big| = \pd{E_n\lr{k}}{k} t$ within the time $t$. 
Taking the maximum with respect to $k$, the wavefront propagates according to
\be
j^{\text{max}}-{i_0} = \max\limits_k \pd{E_n\lr{k}}{k} t = v_{n,\text{max}} t \: .
\label{eq:vmax}
\ee
Apparently, this estimate only yields the ballistic component, while a diffusive component should arise from a dephasing within the linear combination of eigenstates.
This means that if velocities are to be compared with $R\lr{t}$ from DMRG data, a potentially appearing diffusive component has to be filtered out according to Eq.\ (\ref{eq:Rfit}).
Hence, we calculate the maximal velocity $v_{n,\text{max}}$ from the Bethe ansatz and identify it with the velocity $v$ obtained from Eq.\ (\ref{eq:Rfit}). 

Fig.\ \ref{fig:velo} shows $v_{n,\text{max}}$ from various Bethe ansatz excitations displayed as solid lines.
From $n=2$ down to $n \approx 1.4$, we find an almost perfect matching of the velocity of the wavefront with the one of the $k$-$\Lambda$ string (of length 1). 
In this range, the doublon propagates nearly ballistically, and it appears that the effective doublon Hamiltonian Eq.\ (\ref{eq:Heff}) is still approximately valid, the main effect of its interaction term just being a renormalization of the doublon hopping amplitude.
With decreasing filling, however, the presence of more and more singly occupied sites progressively invalidates the effective model, so that the scattering of doublons from single occupancies becomes an interaction process with increasing relevance. 
This is reflected in the increase of the diffusion constant $\ca D$. 

While the additional scattering at intermediate fillings might be expected to contribute to a stronger doublon decay and to hinder doublon propagation, singly occupied sites can also ``accelerate'' a doublon, namely by providing a background on which the propagation does not require a virtual process costing an energy of the order of $U$. 
Indeed, our analysis suggests that scattering at first merely adds a diffusive component to the dynamics, while the ballistic part continues to propagate with the velocity of the bound state. 
It starts to increase and deviate from the velocity of the $k$-$\Lambda$ string below $n \sim 1.4$ (see Fig. \ref{fig:velo}). 

As mentioned above, we expect a crossover between {\em two} ballistic regimes: 
a bound state at $n=2$ (with $v=J$) and a free particle at $n=1$ (with $v=2T$), so that the actual doublon velocity should interpolate between the two cases, requiring an increase toward half filling. 
This effect is obviously not peculiar to one-dimensional systems and has also been observed using nonequilibrium dynamical mean-field theory.\cite{Eckstein_Werner_2014}

%%%%%%%%%%%%%%%%%%%%%%%%%%%%%%%%%%%%%%%%%%%%%%%
%%%%%%%%%%%%%%%%%%%%%%%%%%%%%%%%%%%%%%%%%%%%%%%

However, for times larger than $t \sim 3$ in the filling range $1 < n \lesssim 1.4$, the mixed ballistic-diffusive model breaks down. Evidently, the many-body character of the problem becomes so severe as to disallow an interpretation of the excitation in terms of a single spreading particle. The initial wavefront dies out and the dynamics becomes much slower (see Fig.\ \ref{fig:h_n=1.2}), so that the exact scaling of $R^2\lr{t}$ with $t$ is difficult to read off on the accessible time scale (see the inset of Fig.\ \ref{fig:h_n=1.2}). 
A possible explanation could be that quadruplons ($k$-$\Lambda$ strings of length 2) begin to contribute significantly to the final states. 
This is evidenced by the two-hole spectral function, where the quadruplon peak becomes visible below a filling of $n\approx1.5$. \cite{Rausch_Potthoff_2016} 
In this filling range, the velocity of the quadruplons is very low (see the ``2-string'' line in Fig.\ \ref{fig:velo}), so that they may stay behind as a very slowly expanding core after the doublons have moved away or decayed. The total double occupancy (Fig. \ref{fig:htot}) does not discriminate between empty doublonic and quadruplonic sites, so that this separation can only be seen in the spatially resolved plot.

Note that for $n=1$, the fit of $R^2\lr{t}$ yields a too large velocity of $v\approx2.5T$ and $\mathcal{D}$ becomes negative (see Fig.\ \ref{fig:velo}). 
This is due to the fact that the hole density in the initial state after the application of $d_i$ is spread out across several lattice sites, so that the boundary condition of a strictly localized hole, needed to derive Eq.\ (\ref{eq:rho_xt}), is no longer perfectly fulfilled. 
Alternatively, we can extract the wavefront directly from the data by looking for the point when $\big|\avg{\tilde{n}^h_j}\lr{t}\big|>\epsilon$, where we set $\epsilon=0.05$. 
The respective results for $v$ are displayed in Fig.\ \ref{fig:velo} by red triangles and indeed yield a velocity close to the expected value $2T$ at half filling. 
Similarly, without this artifact, we would expect the diffusion constant to go to zero at $n=1$.

%---------------
\section{Concluding remarks}
\label{sec:summary}
%---------------

Generically, a state with a doublon, i.e., a pair of electron holes at a certain site $i_{0}$ of the lattice, is not an eigenstate of the Hamiltonian, but can be prepared as an initial state in systems of ultracold fermionic atoms loaded in an optical lattice or, in the context of solid-state physics, emerges as the initial state after an essentially local CVV Auger process.
For Hubbard-type systems with strong local Coulomb repulsion $U$, this initial doublon is repulsively bound. 
It forms a compound object which is greatly stabilized by the kinematic constraints set by energy and momentum conservation.
In the present study we have analyzed the fate of the initial doublon: 
its delocalization via propagation through the lattice, its decay on the short time scale before the kinematic constraints become fully active, as well as the resulting decay products.
Generally, for the Hubbard model with an arbitrary finite hole density $2-n$, this requires to tackle a full many-body problem.
For the one-dimensional case, however, some insight is possible thanks to suitable numerical and analytical techniques of DMRG and the Bethe ansatz.
Unfortunately, the presumably extremely long time scale, exponentially increasing with $U$, on which one expects a {\em complete} doublon decay via higher-order scattering processes, is not accessible in this way. 
Still, the physics found on the short and intermediate scale has turned out to be highly relevant and interesting: 

First of all, there is a simple limit, $n=2$, which is fully understood. 
Here, the initial doublon excitation is a linear combination of a bound state and a continuum of scattering states. 
The bound state corresponds to a two-hole compound propagating (in the strong-$U$ limit) with the velocity $v=J=4T^2/U$ through the lattice. 
The continuum of scattering states on the other hand reflects the dispersion of two independently propagating holes.

At half-filling, $n=1$, the {\em creation} of a doublon excitation $c_{i_0\uparrow}c_{i_0\downarrow} \ket{0,N}$ is most unlikely, as this requires a configuration contributing to the ground state $\ket{0,N}$ with a double occupancy at the site $i_{0}$, which must result from a charge fluctuation involving a high energy of the order of $U$.
Once created, however, the doublon is most stable. 
In the strong-$U$ limit which is captured by the $t$-$J$ model with vanishing $J$, it moves with velocity $v=2T$ through the antiferromagnetic background like a free particle, since propagation does not require any virtual process.

For intermediate fillings with $1<n<2$, there are two competing effects:
From the $n=2$ perspective one expects an opening up of decay channels due to the presence of singly occupied sites, and thus a largely increased phase space for scattering, resulting in a clearly enhanced decayed fraction.
Opposed to this, the $n=1$ perspective suggests a stabilization of the doublon, since a background of predominantly singly occupied sites basically undresses the doublon and blocks its decay.
This competition is key to understanding the doublon physics in the different filling regimes.

For fillings above $n \approx 1.4$ (this particular crossover value emerges for the given choice of $U=6$), the dichotomy between a bound state and the scattering-state continuum is kept. 
Contrary to the simple $n=2$ case, however, both have to be understood in terms of Bethe-ansatz eigenstates, as has been explained in the Bethe-ansatz analysis of the two-hole spectral function obtained by DMRG.
While the bound state propagates with the velocity of a $k$-$\Lambda$ string, the scattering states involve two spinons and two holons. 
In particular, a description of the decay products in terms of stable or metastable unbound electron holes is no longer adequate. 
This is an important observation as it implies that, for the one-dimensional case, a {\em sequential} decay process, with a doublon decaying into electron holes which subsequently decay into spinons and holons, must be excluded. 

With decreasing filling, there is an enhanced probability for ground-state configurations with a significant amount of singly occupied sites. 
In the regime $n \gtrsim 1.4$, these lead to an additional scattering of doublons which manifests itself in an additional diffusive component to the doublon propagation. 
However, the bound state is not slowed down since the ballistic component remains (the maximal velocity is not lowered).
The doublon propagation must therefore be seen as a mixed ballistic-diffusive one. 
With respect to the doublon decay we find a decrease of the decayed fraction with decreasing filling.
This indicates that the stabilization effect of singly occupied sites becomes more and more effective.

The doublon physics in the filling regime $1 < n \lesssim 1.4$ is most complicated.
The $U$-dependence of the decay time can no longer be captured by the simple relation $1/t_{\text{min}} \sim U$, which indicates that even on the extremely short time scale, on which the kinematic constraints inhibiting doublon decay are not yet activated, an effective two-site model breaks down and longer-ranged correlations seem to become substantial. 
We also observe that the time dependence of the total double occupancy exhibits a behavior which is qualitatively different from the simple dip-and-plateau-like structure that is found for high fillings.
In fact, $n^h_{\text{tot}}(t)$ shows long-time oscillations, the origin of which is yet unclear. 

Furthermore, doublon propagation in the regime $n \lesssim 1.4$ gets more complicated as well.
With decreasing filling, the maximal velocity deviates more and more from the velocity of the $k$-$\Lambda$ string. 
The doublon becomes increasingly unbound, reflecting the crossover to the $n=1$ case where its propagation is free-particle-like. 
At the same time, however, the mixed ballistic-diffusive propagation model only remains valid for short times $t \lesssim 3$ and breaks down thereafter. 
This is indicated by a qualitatively different light-cone dynamics of the double occupancy radius $R\lr{t}$, which could possibly be explained by the formation of slow quadruplons which are seen in the two-hole excitation spectrum in this filling range. 
One should note that the initial conditions of both the ballistic and the diffusive model involve a single localized particle, while the problem actually attains a many-body character which is impossible to understand in such a simple picture.
\\

\acknowledgments
Financial support of this work through the Deutsche Forschungsgemeinschaft within the Sonderforschungsbereich 925 ``Light induced dynamics and control of correlated quantum systems'' (project B5) is gratefully acknowledged. 

%\bibliography{lit}

\end{document}